\documentclass[12pt,a4paper]{article}
\usepackage[a4paper]{geometry}
\usepackage[utf8]{inputenc}
\usepackage[english]{babel}
\usepackage{amsmath}
\usepackage{amstext}
\usepackage{amscd}
\usepackage{amsfonts}
\usepackage{amssymb}
\usepackage{mathrsfs}

\newtheorem{proposition}{Proposition}

\newtheorem{proof*}{Proof}

\usepackage[pagebackref,draft=false]{hyperref}
\hypersetup{colorlinks,
linkcolor=myrefcolor,
citecolor=mycitecolor,
urlcolor=myurlcolor}

\usepackage[capitalize]{cleveref}
\usepackage{cite}
\usepackage{caption}
\usepackage{etaremune}

\usepackage{xcolor}
\definecolor{myurlcolor}{rgb}{0,0,0.4}
\definecolor{mycitecolor}{rgb}{0,0.5,0}
\definecolor{myrefcolor}{rgb}{0.5,0,0}
\usepackage{graphicx}
\usepackage{tikz}
\usepackage{tikz-cd}
\usetikzlibrary{cd}
\usetikzlibrary{decorations.pathmorphing,shapes}

\usepackage{etoolbox}
\usepackage{enumitem} 
\usepackage[]{latexsym}
\usepackage{braket}
\usepackage{caption}
\usepackage[utf8]{inputenx}
\usepackage[T1]{fontenc}
\usepackage{lmodern}
\usepackage{textcomp}
\usepackage{microtype}
\usepackage{totcount}
\usepackage{blindtext}

\newcommand{\be}{\begin{equation}}
\newcommand{\ee}{\end{equation}}

\title{Dynamical maps and symmetroids}
\date{}

\author{F. M. Ciaglia $^{1,6}$ \href{https://orcid.org/0000-0002-8987-1181}{\includegraphics[scale=0.7]{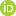}}, F. Di Cosmo$^{1,2,7}$ \href{https://orcid.org/0000-0003-0256-5913}{\includegraphics[scale=0.7]{ORCID.png}}, A. Ibort$^{1,2,8}$ \href{https://orcid.org/0000-0002-0580-5858}{\includegraphics[scale=0.7]{ORCID.png}},\\ G. Marmo$^{3,4,9}$ \href{https://orcid.org/0000-0003-2662-2193}{\includegraphics[scale=0.7]{ORCID.png}} \\
\footnotesize{$^{1}$\textit{Depto. de Matem\'aticas, Univ. Carlos III de Madrid, Legan\'es, Madrid, Spain}} \\
\footnotesize{$^{2}$\textit{ ICMAT, Instituto de Ciencias Matem\'{a}ticas (CSIC-UAM-UC3M-UCM)}}\\
\footnotesize{$^{3}$\textit{ INFN-Sezione di Napoli, Naples, Italy}} \\
\footnotesize{$^{4}$\textit{ Dipartimento di Fisica ``E. Pancini'', Universit\`a di Napoli Federico II,  Naples, Italy}} \\
\footnotesize{$^{6}$\textit{ e-mail: \texttt{fciaglia[at]math.uc3m.es}}} \\
\footnotesize{$^{7}$\textit{ e-mail: \texttt{fcosmo[at]math.uc3m.es}}} \\
\footnotesize{$^{8}$\textit{ e-mail: \texttt{albertoi[at]math.uc3m.es}}} \\
\footnotesize{$^{9}$\textit{ e-mail: \texttt{gmarmo[at]unina.it}}} 
}

\begin{document}

\maketitle

\abstract{Starting from the canonical symmetroid $\mathcal{S}(G)$ associated with a groupoid $G$, the issue of describing dynamical maps in the groupoidal approach to Quantum Mechanics is addressed. After inducing a Haar measure on the canonical symmetroid $\mathcal{S}(G)$, the associated von-Neumann groupoid algebra is constructed. It is shown that the left-regular representation allows to define linear maps on the groupoid-algebra of the groupoid $G$ and given subsets of functions are associated with completely positive maps. Some simple examples are also presented.} 

\section{Introduction}

\textit{In memory of Andrzej Marek Kossakowski, \newline
a noble man, who has been for us a friend, a colleague, a teacher,\newline 
a source of inspiration for his work on open systems}

\vspace{12pt}

The groupoid approach to Quantum Mechanics, also called Schwinger's picture, has been recently proposed by the authors in order to implement Schwinger's vision on the foundations of Quantum mechanics. In a recent series of papers \cite{CIM-2018,CIM-2019b,CIM-2019a,CIM-2019c,CDIM-2020a,CDIM-2020b,CDIMSZ-2021a,CDIMSZ-2021b} several aspects of this formulation have been analysed: observables, statistical interpretation, composition of subsystems, symmetries have been implemented in this picture but many further details are waiting to be discussed in the future. In this paper, we want to add another step to this staircase addressing the description of dynamical maps which need not be algebra automorphisms. 

In the setting of finite dimensional quantum systems, usually described by finite dimensional Hilbert spaces $\mathcal{H}_n=\mathbb{C}^n$, dynamical maps are linear maps $K\,\colon\,M_n(\mathbb{C})\,\rightarrow \, M_n(\mathbb{C})$ which are positive and unital \cite{Sudarshan1961,Asorey2012a,Asorey2012}. An additional requirement, which is usually assumed, is complete positivity, which means that if the system is coupled to an arbitrary ancillary system and the map is trivially extended to the ancilla, the extended map remains positive. Completely positive maps describe, for instance, the Markovian dynamics of open quantum system weakly coupled to the environment, when the initial state of the system is a product state (see for instance \cite{Shabani2009,Shabani2016,UshaDevi2011,Buscemi2014}). Completely positive maps are fundamental ingredients in Quantum Information Theory, where they are called quantum channels and describe admissible quantum operations (see \cite{petz2007quantum,nielsen2001quantum} as general references). However, positive maps which are not completely positive also have been studied, since they emerge in the description of general dynamics of open quantum systems (see for instance \cite{Carteret2008,reference2018,Poincare2017}) and have found many applications as entanglement witnesses \cite{Chruscinski2009,Chen2010}. In the finite dimensional case, Choi's representation theorem characterises completely positive maps using the Choi's matrix \cite{choi1975completely}. Recently, this representation has also been extended to some infinite dimensional situations \cite{Holevo2011,Stormer2015,Friedland2019}. However, the structure of positive maps is much more involved, even in the finite dimensional case  \cite{stormer2012positive,Majewski2020}. 

In this paper we want to present an approach to the description of dynamical maps using the groupoid approach to Quantum Mechanics, in which a quantum system is described by means of a suitable groupoid and its derived structures \cite{CIM-2018,CIM-2019b,CIM-2019a,CIM-2019c}. The departure point is the canonical symmetroid $\tilde{\mathcal{S}}$ associated with a finite quantum system \cite{CDIMS-2021}. Given a finite set $\Omega$ and the pair groupoid $G(\Omega)$, the canonical symmetroid is a groupoid $\tilde{\mathcal{S}}$ built on top of $G(\Omega)$. The elements of $\tilde{\mathcal{S}}$ describe transformations among transitions of $G(\Omega)$ which are written in terms of the action of the groupoid $G(\Omega)$ on itself, by right and left multiplication (see Sec.\ref{canonical symmetroid}). The main result of this work is contained in Sec.\ref{symmetroid-algebra} We introduce the groupoid-algebra associated with the symmetroid $\tilde{S}$ and we will show that completely positive maps can be described by elements of this algebra which satisfy a condition that will be called flat positive semidefiniteness. This result becomes the analogue of the channel-state duality in the Schwinger's picture of Quantum Mechanics. 

The paper is organized as follows: in Sec.2 we discuss previous results on symmetroids and the Schwinger's picture of Quantum Mechanics. The canonical symmetroid $\mathcal{S}(G)$ of a groupoid $G$ is introduced and the group of flat bisections defined. Sec.3 is devoted to the construction of the symmetroid-algebra and an introductory analysis of the relation of its elements and the theory of dynamical maps. A fundamental ingredient in order to define the symmetroid-algebra of the canonical symmetroid $\tilde{\mathcal{S}}$ is the introduction of a Haar measure, as described in \cite{Hahn-1978b,Hahn-1978} and a general procedure to induce a Haar measure on the canonical symmetroid $\mathcal{S}(G)$ starting from a Haar measure on the groupoid $G$ is presented in Sec.\ref{induced-measure} The last section illustrates the relation between dynamical maps and the left regular representation of the symmetroid-algebra. Since the aim of this work is mainly pedagogical, we will focus only on those quantum systems described by finite or countable groupoids in order to avoid all the technical issues related with functional analytic details.       

\section{Symmetries and 2-groupoids}
\subsection{Schwinger's Picture of Quantum Mechanics}\label{SPQM}
In a recent paper \cite{CDIMS-2021} the authors have shown that the concept of symmetry in the groupoid approach to Quantum Mechanics can be encoded into an algebraic structure called 2-groupoid. In particular, due to the partial composition law of a groupoid $G$ there is a 2-groupoid which is canonically associated with any groupoid, which is the 2-groupoid generated by right and left action of the groupoid on itself. Since this mathematical object will play a central role in the second part of the paper, in this section we are going to briefly recall the basic ingredients of Schwinger's picture of Quantum Mechanics and the notions regarding 2-groupoids which will be necessary in what follows. As mentioned in the introduction, since this work is pedagogical in its aim, we will consider only the case of finite (and countable) groupoids leaving the continuum case for future works. 

The departure assumption in Schwinger's picture of Quantum Mechanics consists in associating with every quantum system a pair $(G,\,[\mu])$, where $G\rightrightarrows \Omega$ is a groupoid and $\Omega$ the corresponding space of objects, and $[\mu]$ is an equivalence class of measures on $G$ (two measures are equivalent if they have the same null-measure subsets). Being $G$ countable, the Borel structure on $G$ is trivial, in the sense that every subset is a measurable subset. For the sake of simplicity, let us assume that the groupoid $G$ is connected, i.e., for every pair of objects there is a transition having them as source and target (see below). Elements 
$$
x,y,z\cdots\,\in \Omega
$$ 
will be interpreted as possible outcomes of observables of the system and morphisms 
$$
\alpha\,\colon\,x\,\rightarrow\,y
$$
will be interpreted as ``quantum leaps''. In the rest of the paper we will call them ``transitions''. Two maps,
\begin{equation}
s\,\colon\,G\rightarrow \Omega\,, \qquad t\,\colon\,G\rightarrow \Omega\,,
\end{equation}
called respectively the source and the target, associate every transition to its input and output objects.

A subset of the product space $G^{(2)}\subset G\times G$ is distinguished such that a partial composition exists, the product being denoted $\circ$. In particular, the product $\beta\circ\alpha$ between $\beta,\,\alpha\in G^{(2)}$ is defined whenever the source of $\beta$ coincides with the target of $\alpha$. The axioms satisfied by this composition law are:
\begin{itemize}
\item Associativity;
\item Existence of units: for every outcome $x\in \Omega$ there is a transition, $1_x$ which leaves invariant any other composable transition, i.e., 
\begin{equation}
\alpha \circ 1_x = 1_y \circ \alpha = \alpha\,;
\end{equation}
\item Existence of inverses: for every transition $\alpha\,\colon \,x\rightarrow \, y$, there exists an inverse transition $\alpha^{-1}\,\colon\,y\,\rightarrow\,x$, such that $\alpha\circ \alpha^{-1}=1_y$ and $\alpha^{-1}\circ\alpha = 1_x$ (this property is instrumental for implementing Feynman's principle of ``microscopic reversibility''\cite{Feynman-2005}). 
\end{itemize}

As proven by Hahn \cite{Hahn-1978b}, in the class $[\mu]$ there is a Haar measure\footnote{\begin{scriptsize} The name comes from the fact that, when the groupoid $G$ is actually a locally compact topological group, then $\mu$ reduces to the standard Haar measure.\end{scriptsize}}, which with an abuse of notation we will call $\mu$, which satisfies the following properties:
\begin{itemize}
\item There exists a family of measures $\nu^x$, with $x\in \Omega$, with support on $G^x := t^{-1}(x)$ such that for any measurable set $E$
\begin{equation}\label{eq.2.3}
\mu(E) = \int_{\Omega} \nu^x(E) \,\mathrm{d}\mu_{\Omega}(x)\,,
\end{equation}
where $\mu_{\Omega}= t_{\star}\mu = s_{\star}\mu$, is a measure on the space of objects $\Omega$; 
\item Let $\tau\,\colon\,G\,\rightarrow\,G$ be the inversion map associating to any transition $\alpha\in G$ its inverse $\alpha^{-1}$, the measure $\mu^{-1} = \tau_{\ast} \mu$ is absolutely continuous with respect to $\mu$ and its Radon-Nikodym derivative
\begin{equation}
\frac{\mathrm{d}\mu^{-1}}{\mathrm{d}\mu} = \delta^{-1}
\end{equation} 
is a groupoid homomorphism $\delta\,\colon\,G\,\rightarrow \,\mathbb{R}_+$ called the modular function;
\item The family of measures $\nu^x$ is left-invariant, in the sense that, if $\gamma\,\colon\,x\,\rightarrow\,y$ is a transition and $L_{\gamma}\,\colon\,G^x\,\rightarrow \,G^y$ the left action $L_{\gamma}(\alpha)= \gamma \circ \alpha$, then 
\begin{equation}\label{equivariance_G}
(L_{\gamma})_{\ast} \nu^x = \nu^y\,,
\end{equation} 
a property which extend to the category of groupoids the concept of left-invariant measure for groups. 
\end{itemize} 
From the above list of properties one can also derive the existence of another family of measures $\nu_x$, with $x\in \Omega$ and support contained in $G_x:=s^{-1}(x)$ such that 
\begin{equation}\label{inverse_nu}
\tau_{\ast}(\nu^x) = \delta^{-1}\nu_x \,.
\end{equation} 
Contrarily to the family $\nu^x$, the family $\nu_x$ is right-invariant, i.e., $(R_{\gamma})_{\ast} \nu_y = \nu_x$. 

The space, $C_0(G)$ of measurable (and continuous for this discrete case) complex valued functions on $G$ with compact support, can be equipped with an involution operator, say $\dagger$, and a convolution product, $\star_{\nu}$. Given two function $f,\,g\in C_0(G)$ the convolution product is defined as follows:
\begin{equation}
f\star g (\alpha) = \sum_{\beta \in G^{t(\alpha)}} f(\beta) g(\beta^{-1}\circ\alpha) \nu^{t(\alpha)}(\beta)\,, 
\end{equation}
whereas the involution is given by
\begin{equation}
f^{\ast}(\alpha) = \delta^{-1}(\alpha)f(\alpha^{-1})\,.
\end{equation}
Then, one can consider the space $\mathcal{L}^2(G, \mu)=:\mathcal{H}_0$ of square-integrable functions on $G$ with respect to the measure $\mu$, and $C_0(G)$ acts on it as follows: given $f\in C_0(G)$ and $\psi \in \mathcal{H}_0$ the representation $\pi\,\colon\, C_0(G)\,\rightarrow \, B(\mathcal{H}_0)$ is defined as:
\begin{equation}
\pi(f)\psi = f\star \psi\,.
\end{equation}
The completion with respect to the weak-operator topology provides us with a von-Neumann algebra $\mathcal{V}(G)$, which we call the von-Neumann groupoid-algebra of the groupoid $G$.  
The real elements of this von-Neumann algebra represent the observables of the system under investigation, whereas states of the system are represented by states of the von-Neumann algebra $\mathcal{V}(G)$, which means positive, linear and normalised functionals on $\mathcal{V}(G)$. In the discrete countable case, every state $\rho$ determines a function of positive-type $\varphi_{\rho} \in C^*(G)$:
\begin{equation}
\rho (\delta_{\alpha}) = \varphi_{\rho}(\alpha)\,,
\end{equation}
where $\delta_{\alpha}$ is the function on $G$ which takes the value $1$ only at the transition $\alpha$. A positive definite function on a groupoid $G$ is defined as follows: For any $N\in \mathbb{N}$, and for every collection of complex numbers $\left\lbrace \xi_j \right\rbrace_{j=1,2,\cdots, N}$ and transitions $\left\lbrace  \alpha_j\right\rbrace$ the following holds
\begin{equation}\label{positive_definite_function}
\sum_{s(\alpha_j)=s(\alpha_k)}\xi_j \overline{\xi}_k \varphi_{\rho}(\alpha_j\circ\alpha_k^{-1}) \geq 0 \,.
\end{equation}

A natural relation between states in the Schwinger's picture of Quantum Mechanics and quantum measures (see \cite{sorkin1994quantum} for the definition of quantum measures) has been shown in \cite{CIM-2019c}. Additionally, in the countable case one can describe normal state $\omega_{\xi}$ \cite{Blackadar-2006}, in terms of sequences of functions $\xi_n$ with $\xi_n\in \mathcal{H}_0$ for every $n\in \mathbb{N}$ such that 
\begin{equation}
\sum_{n=1}^{\infty} \parallel \xi_n \parallel_{\mathcal{H}_0}^2 = 1\,.
\end{equation} 
Then, given an element $a\in \mathcal{V}(G)$, the evaluation of the state $\omega_{\xi}$ on $a$ is written as follows:
\begin{equation}
\omega(a) = \sum_{n=1}^{\infty} \left\langle \xi_n ,\, a\xi_n \right\rangle_{\mathcal{H}_0}\,,
\end{equation}
where $\left\langle \cdot , \, \cdot \right\rangle_{\mathcal{H}_0}$ denotes the scalar product of the Hilbert space $\mathcal{H}_0$. 

\subsection{The canonical symmetroid $\mathcal{S}(G)$}\label{canonical symmetroid}
The concept of symmetry of a quantum system can be encoded in Schwinger's picture of Quantum Mechanics associating with a symmetry of a quantum system a 2-groupoid, which in this context has been called symmetroid \cite{CDIMS-2021}. In particular, since a groupoid $G$ acts on itself by left and right multiplication, given $G$ one can construct a canonical symmetroid, called $\mathcal{S}(G)$, and this construction will be the subject of this section.

Let $G\rightrightarrows \Omega$ be a finite (or countable) groupoid on the discrete set $\Omega$. An element of a 2-groupoid is a morphism between transitions of the groupoid $G$, which we will call a transformation. The canonical little symmetroid $\mathcal{S}_0(G)\rightrightarrows G\rightrightarrows \Omega $ is generated by the transformations $(L_{\alpha},\beta)$ and $(R_{\gamma},\beta)$ with $\alpha\in \mathrm{Aut}(t(\beta))$ and $\gamma \in \mathrm{Aut}(s(\beta))$ and targets given by:
\begin{equation}
L_{\alpha}(\beta) = \alpha\circ \beta, \quad R_{\gamma}(\beta) = \beta\circ\gamma^{-1}\,.
\end{equation}
Therefore, a generic element $\Gamma\in \mathcal{S}_0(G)$ will be denoted by a triple $\Gamma\equiv (\alpha,\beta,\gamma)$ such that the product $\alpha\circ\beta\circ\gamma^{-1}$ belongs to $G({s(\beta)},{t(\beta)})$, i.e., the subset of transitions between $s(\beta)$ and $t(\beta)$. On $\mathcal{S}_0(G)$ we define a pair of maps $s_1\,\colon\,\mathcal{S}_0(G)\,\rightarrow\,G$ and $t_1\,\colon\,\mathcal{S}_0(G)\rightarrow\,G$, called respectively the 2-source and the 2-target maps, as follows:
\begin{equation}
s_1(\Gamma) = \beta,\quad t_1(\Gamma)= \alpha\circ\beta\circ\gamma^{-1}\,.
\end{equation}  
Then the subset $\mathcal{S}_0^{(2)}$ of pairs $(\Gamma_1,\Gamma_2)\in \mathcal{S}_0(G)\times\mathcal{S}_0(G)$ such that $t_1(\Gamma_1)=s_1(\Gamma_2)$ form a subset of composable transformations with respect to the vertical composition $\circ_V\,\colon\,\mathcal{S}^{(2)}\,\rightarrow\,\mathcal{S}_0(G)$:
\begin{equation}
\Gamma_2\circ_V\Gamma_1 = (\alpha_2,\beta_2,\gamma_2 )\circ_V(\alpha_1,\beta_1,\gamma_1) = (\alpha_2\circ\alpha_1,\beta_1,\gamma_2\circ\gamma_1)\,,
\end{equation}
where the condition $t_1(\Gamma_1)=s_1(\Gamma_2)$ means $\alpha_1\circ\beta_1\circ\gamma_1^{-1} = \beta_2$. It is a matter of straightforward computations to prove that the product $\circ_V$ satisfies the following properties:
\begin{itemize}
\item It is associative;
\item For every $\beta\in G$ there is a transformation $\mathbf{1}_{\beta}\,\colon\,\beta\,\rightarrow\,\beta$, called units;
\item For every $\Gamma = (\alpha,\beta,\gamma)\in \mathcal{S}_0(G)$ there is an inverse transformation $\Gamma^{-1}=(\alpha^{-1},\alpha\circ\beta\circ\gamma^{-1},\gamma^{-1})$ such that $\Gamma\circ_V\Gamma^{-1}=\mathbf{1}_{t_1(\Gamma)}$ and $\Gamma^{-1}\circ_V\Gamma = \mathbf{1}_{s_1(\Gamma)}$.
\end{itemize}  
Therefore the set $\mathcal{S}_0(G)\rightrightarrows G$ endowed with the vertical composition is a groupoid over the groupoid $G$. Since the set $G$ is a groupoid, a second composition can be defined on the set $\mathcal{S}_0(G)$, called horizontal composition. Indeed, given two transformations $\Gamma_1\,\colon\,\beta_1\,\rightarrow \, \beta'_1$ and $\Gamma_2\,\colon\,\beta_2\,\rightarrow\,\beta'_2$ such that the pairs $(\beta_2,\beta_1)$ and $(\beta'_2,\beta'_1)$ are both composable, the horizontal composition $\Gamma_2\circ_H\Gamma_1$ is the transformation from the composition $\beta_2\circ\beta_1$ to $\beta'_2\circ\beta'_1$ which can be written as
\begin{equation}
\Gamma_{2}\circ_H \Gamma_1 = (t_1(\Gamma_2)\circ s_1(\Gamma_2)^{-1}, s_1(\Gamma_2)\circ s_1(\Gamma_1), t_1(\Gamma_1)^{-1}\circ s_1(\Gamma_1))\,.
\end{equation}  
It is possible to prove via long but straightforward computations that the horizontal composition satisfies the following properties:
\begin{itemize}
\item It is associative;
\item For every $x\in\Omega$ there is a horizontal unit transformation $\mathbb{I}_x=(1_x,1_x,1_x)$ which leaves the unit $1_x$ invariant;
\item For every transformation $\Gamma = (\alpha, \beta, \gamma)$ there is a horizontal inverse $\Gamma^{-H} = (\gamma^{-1}, \beta^{-1}, \alpha^{-1})$ such that 
\begin{equation}
\Gamma^{-H}\circ_H \Gamma = \mathbb{I}_{s(\beta)},\quad \Gamma\circ_H \Gamma^{-H} = \mathbb{I}_{t(\beta)}\,.
\end{equation} 
\end{itemize}
The horizontal composition, however, does not define a groupoid structure over $G$ with respect to the 2-source and 2-target maps, since a pair of transitions $(\Gamma_1, \Gamma_2)$ can be composed horizontally even if $s_1(\Gamma_2)\neq t_1(\Gamma_1)$. Finally, it can also been shown that the vertical and horizontal composition satisfy a compatibility condition known as the exchange identity:
\begin{equation}
(\Gamma'_2\circ_H\Gamma'_1)\circ_V(\Gamma_2\circ_H\Gamma_1) = (\Gamma'_2\circ_V\Gamma_2)\circ_H(\Gamma'_1\circ_V\Gamma_1)\,.
\end{equation}

The canonical little symmetroid introduced above can, now, be extended. Let us consider the transformation $\Gamma=(\alpha, \beta, \gamma)$ in $\mathcal{S}_0(G)$. Let us replace this pair of transitions with transitions $(\tilde{\alpha},\,\tilde{\gamma})$ such that both $(\tilde{\alpha},\,\beta)$ and $(\beta,\,\tilde{\gamma}^{-1})$ are both composable pair, and call it again ${\Gamma}$, with an abuse of notation. Generalizing in the obvious way the previously defined $2$-source, $2$-target and vertical composition $\circ_V$, we obtain a groupoid $\mathcal{S}(G)$ over $G$, called the canonical symmetroid of the groupoid $G$. Analogously, one can define the horizontal composition between transformations $\Gamma_1\,\colon\,\beta_1\,\rightarrow \, \beta'_1=(\alpha_1\circ \beta_1\circ \gamma_1^{-1}) $ and $\Gamma_2\,\colon\,\beta_2=(\alpha_2\circ \beta_2\circ \gamma_2^{-1})\,\rightarrow\,\beta'_2$ such that the pairs $(\beta_2,\beta_1)$ and $(\beta'_2,\beta'_1)$ are both composable. The horizontal composition $\Gamma_2\circ_H\Gamma_1$ is the transformation from the composition $\beta_2\circ\beta_1$ to $\beta'_2\circ\beta'_1$ which can be written as
\begin{equation}
\Gamma_{2}\circ_H \Gamma_1 = (t_1(\Gamma_2)\circ \alpha_1 \circ s_1(\Gamma_2)^{-1}, s_1(\Gamma_2)\circ s_1(\Gamma_1), \gamma )\,.
\end{equation}  
This composition satisfies the following properties:
\begin{itemize}
\item It is associative;
\item For every pair of objects $(x,y)$ there is a horizontal unit transformation $\mathbb{I}_{yx}:=(\alpha_{yx},1_x,\alpha_{yx})$ which sends the unit transition $1_x$ to the unit transition $1_y$ and it is defined via the choice of a transition $\alpha_{yx}\in G(x,y)$. Then, we have that
\begin{equation}
\mathbb{I}_{t(t_1(\Gamma)),t(s_1(\Gamma))}\circ_H \Gamma = \Gamma = \Gamma \circ_H \mathbb{I}_{s(t_1(\Gamma)),s(s_1(\Gamma))}\,.   
\end{equation}
\item Once chosen the horizontal unit transformations, then for every transformation $\Gamma = (\alpha,\,\beta,\, \gamma)$, there is an inverse transformation $\Gamma^{-H}= (\alpha',\, \beta^{-1},\, \gamma' )$ such that 
\begin{equation}\label{Eq.2.22}
\Gamma^{-H}\circ_{H}\Gamma = \mathbb{I}_{s(s_1(\Gamma)), s(t_1(\Gamma))},\quad \Gamma\circ_{H}\Gamma^{-H} \approx \mathbb{I}_{t(s_1(\Gamma)), t(t_1(\Gamma))}\,,
\end{equation}
where the symbol $\approx$ denotes the equivalence up to transformation in $\mathcal{S}_0(G)$. From Eq.(\ref{Eq.2.22}) we derive that $\gamma' = \alpha_{t(s_1(\Gamma)), t(t_1(\Gamma))}$ and $\alpha' = t_1(\Gamma)^{-1}\circ \alpha_{t(s_1(\Gamma)), t(t_1(\Gamma))} \circ \beta$.  
\end{itemize}
The canonical little symmetroid $\mathcal{S}_0(G)$ is a normal subgroupoid (with respect to the vertical composition) of the symmetroid $\mathcal{S}(G)$, since the isotropy subgroupoid of the groupoid $\mathcal{S}_0(G)$, which is made up of identity transformations only,  is left invariant by conjugation with respect to transformation in $\mathcal{S}(G)$, i.e., given $\mathbf{1}_{\beta}\in \mathcal{S}_0(G)$ and $\Gamma \in \mathcal{S}(G)$ such that $\Gamma\,\colon\,\beta \,\rightarrow\,\beta'$ we have that 
\begin{equation}
\Gamma\circ_V \mathbf{1}_{\beta}\circ_V \Gamma^{-1} = \mathbf{1}_{\beta'}\,.
\end{equation}
Therefore, the quotient set 
\begin{equation}
\tilde{\mathcal{S}} = \mathcal{S}(G)/ \mathcal{S}_0(G)
\end{equation}
is a groupoid itself\cite{Ibort2019}. In particular it holds that
\begin{equation}
\tilde{\mathcal{S}} \rightrightarrows G(\Omega) \rightrightarrows \Omega\,,
\end{equation}
where $G(\Omega)$ denotes the pair-groupoid over the set $\Omega$, is a proper groupoid. Indeed, any equivalence class of transformation $[\Gamma]= [(\alpha, \beta, \gamma)]$ can be identified with a couple of pairs of points in $\Omega$, i.e., $[\Gamma] \cong  ( (t(t_1(\Gamma)), t(s_1(\Gamma))), (s(s_1(\Gamma)), s(t_1(\Gamma))) )$, where $\Gamma$ is any transformation in the equivalence class. Then, we can define the 2-source map, $\tilde{s}_1\,\colon\, \tilde{\mathcal{S}}\,\rightarrow \, G(\Omega)$ and 2-target map $\tilde{t}_1\,\colon\, \tilde{\mathcal{S}}\,\rightarrow \, G(\Omega)$ as follows
\begin{equation}
\tilde{s}_1([\Gamma]) = (t(s_1(\Gamma)), s(s_1(\Gamma))) ,\quad \tilde{t}_1([\Gamma]) = (t(t_1(\Gamma)), s(t_1(\Gamma)))\,,
\end{equation}    
such that the vertical and horizontal composition:
\begin{eqnarray}
&((z,y),(x,w))\tilde{\circ}_V ((z',y'),(x',w')) = \delta(y,z')\delta(x,w') ((z,x'),(y',w)) \\
&((z,y),(x,w))\tilde{\circ}_H ((z',y'),(x',w')) = \delta(w,z')\delta(x,y') ((z,y),(x,w))\,,
\end{eqnarray}
satisfy the properties of a symmetroid. In particular the vertical unit transformations are 
\begin{equation}
\mathbf{1}_{y,x} = ((y,y),(x,x))\,, 
\end{equation}
while the horizontal unit transformations are 
\begin{equation}
\mathbb{I}_{y,x} = ((y,x),(x,y))
\end{equation}
\subsection{Bisections, flat bisections and groups of symmetries}\label{sec:flat_bisections}
Given the canonical symmetroid $\mathcal{S}(G)\rightrightarrows G \rightrightarrows \Omega$ a bisection \cite{Mackenzie-2005,CDIMS-2021} $\mathbf{b}\subset \mathcal{S}(G)$ is a subset of the canonical symmetroid such that the 2-source and the 2-target maps restricted to $\mathbf{b}$ are bijections. Therefore, given a bisection $\mathbf{b}$ there is a pair of sections $(\mathbf{b}_{s_1},\mathbf{b}_{t_1})$ of the fibered space $\mathcal{S}(G)\rightrightarrows G$, i.e., a pair of maps $\mathbf{b}_{s_1}\,\colon\,G\,\rightarrow\, \mathcal{S}(G)$ and $\mathbf{b}_{t_1}\,\colon\,G\,\rightarrow\, \mathcal{S}(G)$ such that 
\begin{equation}
s_1\circ \mathbf{b}_{s_1} = \mathrm{id}_G,\quad t_1\circ \mathbf{b}_{t_1} = \mathrm{id}_G\,,
\end{equation}  
and one can describe the bisection $\mathbf{b}$ using one of these sections. It can be proven that the set $\mathfrak{S}$ of all bisections of $\mathcal{S}(G)$ is a group under the multiplication $\mathbf{b}'\bullet \mathbf{b}$, whose associated section is defined as follows
\begin{equation}
(\mathbf{b}'\bullet \mathbf{b})_{s_1}(\beta) = \mathbf{b}'_{s_1}(\varphi_{b}(\beta))\circ \mathbf{b}_{s_1}(\beta)\,.  
\end{equation} 
In the above expression, the map
\begin{equation}\label{functor-bisection}
\varphi_{b}(\beta) = t_1((\mathbf{b}_1)_{s_1}(\alpha))
\end{equation}
is an invertible map of the set $G$. 
Given a bisection $\mathbf{b}\subset \mathcal{S}(G)$ the corresponding section $\mathbf{b}_{s_1}$ satisfies the property
\begin{equation}\label{weak-horizontally}
\mathbf{b}_{s_1}(\beta'\circ\beta)\approx \mathbf{b}_{s_1}(\beta')\circ_H\mathbf{b}_{s_1}(\beta)\,,
\end{equation}
where the symbol $\approx$ has been introduced in the previous section. The bisections $\mathbf{b}$ for which the condtion Eq.(\ref{weak-horizontally}) is actually an exact equivalence are called flat bisections and they form a subgroup $\mathfrak{S}^{\flat}\subset \mathfrak{S}$.  

Given a flat bisection the map $\varphi_b\,\colon\, G \,\rightarrow\,G$ defined in Eq.(\ref{functor-bisection}) is an automorphism of the groupoid, i.e., an invertible endofunctor on $G\rightrightarrows \Omega$. Any of these functors is written in terms of a bisection $b\subset G$ as follows
\begin{equation}\label{flat_functor}
\varphi_b(\beta) = b_s(y)\circ \beta \circ b_s(x)^{-1}\,.
\end{equation} 
Indeed, in this way units are invertibly mapped into units and the composition law is preserved. The subgroup of the flat bisection $\mathfrak{S}^{\flat}$ is called the canonical group of symmetries of the groupoid $G$. 

\section{Symmetroid-Algebras and dynamical maps}\label{symmetroid-algebra}
\subsection{The induced measure on the symmetroid $\mathcal{S}(G)$}\label{induced-measure}
In this section we will show that given a measure groupoid $(G,[\mu])$ satisfying the conditions presented in Sec.2.1, it is possible to induce a measure on the canonical symmetroid $\mathcal{S}(G)$. We will use this measure to build up the groupoid-algebra of the symmetroid with respect to the vertical composition law and we will call this algebra the canonical symmetroid-algebra of the groupoid $G$. We will show that this algebra can be used to describe dynamical maps on the space of (normal) states of the von-Neumann groupoid-algebra $\mathcal{V}(G)$. In the rest of this section we will focus on finite dimensional quantum systems, so that we can avoid to deal with mathematical technicalities which, however, will be dealt with in forthcoming works. Let us remark, eventually, that the method which will be adopted to induce a measure on the symmetroid can be extended to the infinite dimensional setting, both countable and not. In particular, an useful mathematical tool to get this result is the notion of transverse function and transverse measure introduced by Connes in his noncommutative theory of integration \cite{non-commutative-1978,kastler1982connes}.  

Let us consider a finite connected groupoid $G\rightrightarrows \Omega$ with a Haar measure $\mu$ and the canonical symmetroid $\mathcal{S}(G)\rightrightarrows G \rightrightarrows \Omega$. As illustrated in Sec.\ref{canonical symmetroid} the canonical symmetroid $\mathcal{S}(G)$ can be described by triples of transitions $\Gamma = (\alpha, \beta, \gamma)\in \mathcal{S}(G)$ of the groupoid $G$ such that $(\alpha,\beta)$ and $(\beta,\gamma^{-1})$ are both composable. Then, we consider on $\mathcal{S}(G)$ the algebra of measurable sets generated by the multirectangle $A\times B\times C$, with $A,\,B,\,C$ subsets of $G$. In particular, in the finite (and countable) case the structure we obtain is still trivial, where every point is a measurable set. A system of left Haar measures can be induced on $\mathcal{S}(G)$ as follows. Let $\mathcal{S}^{\beta}$ be the subset 
\begin{equation}
\mathcal{S}^{\beta} = t_1^{-1}(\beta) = \left\lbrace  \Gamma = (\alpha, \beta', \gamma)\in\mathcal{S}(G) | \alpha\circ \beta'\circ \gamma^{-1}= \beta\right\rbrace\,.
\end{equation}  
One can identify $\mathcal{S}^{\beta}$ with the space $G^{t(\beta)}\times G^{s(\beta)}$ by defining the following bijective map $\Psi_{(\beta)}\,\colon\,\mathcal{S}^{\beta}\,\rightarrow\,G^{t(\beta)}\times G^{s(\beta)}$:
\begin{eqnarray}
& \Psi_{(\beta)}(\alpha, \beta, \gamma) = (\alpha, \gamma) \\ \label{equivalence:eq.1}
& \Psi_{(\beta)}^{-1}(\alpha, \gamma) = (\alpha, \beta', \gamma)\, \label{equivalence:eq.2}
\end{eqnarray}
where $\beta'$ in Eq.(\ref{equivalence:eq.2}) is the unique $\beta'\in G$ satisfying the condition
\begin{equation}
\beta = \alpha\circ \beta' \circ \gamma^{-1}\,.
\end{equation}
Since the space $G^{t(\beta)}\times G^{s(\beta)}$ possesses the measure $\nu^{t(\beta)}\times \nu^{s(\beta)}$ inherited by the Haar measure on $G$, see Eq.(\ref{eq.2.3}), we can define on $\mathcal{S}^{\beta}$ the following measure
\begin{equation}
\nu_{(2)}^{\beta}(A) = (\Psi^{-1}_{(\beta)})_{\ast}(\nu^{t(\beta)}\times \nu^{s(\beta)})(A) = \nu^{t(\beta)}\times \nu^{s(\beta)}(\Psi_{(\alpha)}(A))\,,
\end{equation}
where $A\subset \mathcal{S}^{\beta}$. Then we can define the measure $\mu_{(2)}$ on $\mathcal{S}(G)$ as follows
\begin{equation}
\mu_{(2)}(E) = \int_{G} \nu_{(2)}^{\beta}(E)\,\mathrm{d}\mu (G)\,,
\end{equation}
where $E$ is a measurable subset of $\mathcal{S}(G)$. Let us prove, now, that the family of measures $\nu_{(2)}^{\beta}$ is equivariant under the left action of the symmetroid on itself.

Let $\Gamma\,\colon\,\beta_1\,\rightarrow \,\beta_2 $ be a transformation in the symmetroid $\mathcal{S}(G)$ and $L_{\Gamma}\,\colon\,\mathcal{S}^{\beta_1}\,\rightarrow\,\mathcal{S}^{\beta_2}$ the map corresponding to the left multiplication by $\Gamma$. We want to prove that 
\begin{equation}\label{equivariance}
(L_{\Gamma})_{\ast}\nu_{(2)}^{s_1(\Gamma)} = \nu_{(2)}^{t_1(\Gamma)}\,.
\end{equation}
To get this, let us start from the left-hand-side of the above Eq.(\ref{equivariance}) and from a subset $A\subset \mathcal{S}^{t_1(\Gamma)}$. Then, we have that the following chain of equalities holds:
\begin{eqnarray}
& \nonumber ((L_{\Gamma})_{\ast}\nu_{(2)}^{s_1(\Gamma)}) (A) = \nu_{(2)}^{s_1(\Gamma)} (L_{\Gamma^{-1}}(A)) = \nu^{t(s_1(\Gamma))}\times \nu^{s(s_1(\Gamma))}(\Psi_{s_1(\Gamma)}^{-1}(L_{\Gamma^{-1}}(A))) = \\ \nonumber
& \nu^{s(\alpha)}\times \nu^{s(\gamma)}((L_{\alpha^{-1}}\otimes L_{\gamma^{-1}})(\Psi_{t_1(\Gamma)}^{-1}(A))) = (L_{\alpha}\otimes L_{\gamma})_{\ast}(\nu^{s(\alpha)}\times \nu^{s(\gamma)})(\Psi_{t_1(\Gamma)}^{-1}(A)) = \\ 
& =  \nu^{t(\alpha)}\times \nu^{t(\gamma)}(\Psi_{t_1(\Gamma)}^{-1}(A)) = \nu^{t(t_1(\Gamma))}\times \nu^{s(t_1(\Gamma))}(\Psi_{t_1(\Gamma)}^{-1}(A)) = \nu_{(2)}^{t_1(\Gamma)}(A)\,.
\end{eqnarray}
In other words, the equivariance of the Haar measure $\mu$ on $G$ induces the equivariance of the family of measures $\nu_{(2)}^{\beta}$ on $\mathcal{S}(G)$. A system of measures which is right invariant can be analogously constructed by replacing the set $\mathcal{S}^{\beta}$ with the set $\mathcal{S}_{\beta}= s_1^{-1}(\beta)$ and identifying this set with the set $G_{t(\beta)}\times G_{s(\beta)}$ via a map $\tilde{\Psi}_{(\beta)}$. Then, we define the family of measures $\nu_{(2)\beta}$ on $\mathcal{S}_{\beta}$ as 
\begin{equation}
\nu_{(2)\beta} = (\tilde{\Psi}^{-1}_{(\beta)})_{\ast}(\nu_{t(\beta)}\times \nu_{s(\beta)})\,.
\end{equation} 
Let us now show that this measure is also absolutely continuous with respect to its inverse. Indeed, let $f\,\colon\,\mathcal{S}(G)\,\rightarrow\,\mathbb{C}$. Then, we can evaluate the following integral (in the finite or countable case the integral should actually be a sum or a series but we will adopt the integral symbol since the result holds in the case of measurable spaces) 
\begin{eqnarray}
&\int_{\mathcal{S}(G)}\mathrm{d}\mu_{(2)}(\Gamma)f(\Gamma^{-1})= \int_G\int_{G^{t(\beta)}\times G^{s(\beta)}}\mathrm{d}\mu(\beta)\mathrm{d}\nu^{t(\beta)}(\alpha)\mathrm{d}\nu^{s(\beta)}(\gamma)f(\alpha^{-1}, \gamma^{-1},\alpha^{-1}\circ \beta \circ\gamma)= \nonumber \\
& = \int_G\int_{G_{t(\beta)}\times G_{s(\beta)}}\mathrm{d}\mu(\beta)\mathrm{d}\nu_{t(\beta)}(\alpha)\mathrm{d}\nu_{s(\beta)}(\gamma)\delta^{-1}(\alpha)\delta^{-1}(\gamma)f(\alpha, \gamma,\alpha\circ \beta \circ\gamma^{-1}) = \nonumber \\
& = \int_{\Omega}\int_{G_{y}}\int_{G_y\times G_{t(\beta)}}\mathrm{d}\mu_{\Omega}(y)\mathrm{d}\nu_y{\beta}\mathrm{d}\nu_{t(\beta)}(\alpha)\mathrm{d}\nu_{s(\beta)}(\gamma)\delta^{-1}(\alpha)\delta^{-1}(\gamma)f(\alpha, \gamma,\alpha\circ \beta \circ\gamma^{-1}) = \nonumber \\
& =\int_{\Omega}\int_{G_{y}}\int_{G_{s(\gamma)}\times G_{t(\beta)}}\mathrm{d}\mu_{\Omega}(y)\mathrm{d}\nu_y({\gamma})\mathrm{d}\nu_{s(\gamma)}(\beta)\mathrm{d}\nu_{t(\beta)}(\alpha)\delta^{-1}(\alpha)\delta^{-1}(\gamma)f(\alpha, \gamma,\alpha\circ \beta \circ\gamma^{-1}) = \nonumber \\
& =\int_G\int_{G_{t(\gamma)}\times G_{t(\beta)}}\mathrm{d}\mu({\gamma})\mathrm{d}\nu_{t(\gamma)}(\beta)\mathrm{d}\nu_{t(\beta)}(\alpha)\delta^{-1}(\alpha)\delta^{-1}(\gamma)f(\alpha, \gamma,\alpha\circ \beta) = \nonumber \\
& =\int_{\Omega}\int_{G^{y}}\int_{G_y\times G^{t(\beta)}}\mathrm{d}\mu_{\Omega}(y)\mathrm{d}\nu^y({\gamma})\mathrm{d}\nu_y(\beta)\mathrm{d}\nu_{t(\beta)}(\alpha)\delta^{-1}(\alpha)\delta^{-1}(\gamma)f(\alpha, \gamma,\alpha\circ \beta) = \nonumber \\
& =\int_{\Omega}\int_{G^{y}}\int_{G_{y}\times G^{s(\beta)}}\mathrm{d}\mu_{\Omega}(y)\mathrm{d}\nu^y({\beta})\mathrm{d}\nu_{y}(\alpha)\mathrm{d}\nu^{s(\beta)}(\gamma)\delta^{-1}(\alpha)\delta^{-1}(\gamma)f(\alpha, \gamma,\alpha\circ \beta) = \nonumber \\
& =\int_G\int_{G^{s(\alpha)}\times G^{s(\beta)}}\mathrm{d}{\mu}(\alpha)\mathrm{d}\nu^{s(\alpha)}({\beta})\mathrm{d}\nu^{s(\beta)}(\gamma)\delta^{-1}(\alpha)\delta^{-1}(\gamma)f(\alpha, \gamma,\alpha\circ \beta) = \nonumber \\
& =\int_G\int_{G^{t(\alpha)}\times G^{s(\beta)}}\mathrm{d}{\mu}(\alpha)\mathrm{d}\nu^{t(\alpha)}({\beta})\mathrm{d}\nu^{s(\beta)}(\gamma)\delta^{-1}(\alpha)\delta^{-1}(\gamma)f(\alpha, \gamma,\beta) = \nonumber \\
& =\int_{\Omega}\int_{G^y}\int_{G^{y)}\times G^{s(\beta)}}\mathrm{d}\mu_{\Omega}(y)\mathrm{d}\nu^{y}(\alpha)\mathrm{d}\nu^{y}({\beta})\mathrm{d}\nu^{s(\beta)}(\gamma)\delta^{-1}(\alpha)\delta^{-1}(\gamma)f(\alpha, \gamma, \beta) = \nonumber \\
& = \int_G\int_{G^{s(\beta)}\times G^{t(\beta)}}\mathrm{d}\mu(\beta)\mathrm{d}\nu^{s(\beta)}(\gamma)\mathrm{d}\nu^{t(\beta)}(\alpha)\delta^{-1}(\alpha)\delta^{-1}(\gamma)f(\alpha, \gamma, \beta) = \nonumber \\
& =\int_{\mathcal{S}(G)}\mathrm{d}\mu_{(2)}(\Gamma)\Delta_{(2)}(\Gamma)^{-1}f(\Gamma)
\end{eqnarray}
and we obtain that the modular function $\Delta_{(2)}$ associated to the measure $\mu_{(2)}$ on the canonical symmetroid $\mathcal{S}(G)$ is derived from the modular function $\delta$ associated to the Haar measure on $G$ and it is expressed as
\begin{equation}
\Delta_{(2)}(\Gamma) = \Delta_{(2)}(\alpha,\beta, \gamma) = \delta(\alpha)\delta(\gamma)\,.
\end{equation}
Let us notice that in the previous chain of equalities we have used the properties of the family of measures $\nu^{y}$ expressed in Eq.(\ref{equivariance_G}) and Eq.(\ref{inverse_nu}), and the disintegration of the Haar measure $\mu$ on $G$. Moreover, since $\delta$ is a homomorphism of the groupoid $G$, the modular function $\Delta_{(2)}$ is a homomorphism of the symmetroid with respect to the vertical composition.

\subsection{The von-Neumann symmetroid-algebra}
Once we have the Haar measure $\mu_{(2)}$ on the symetroid $\mathcal{S}(G)$ we can associate with it a von-Neumann algebra, which we will call the von-Neumann symmetroid-algebra. In order to avoid technicalities which could obscure the goal of this paper, from now on we are going to adapt the notation to the case of finite or countable groupoids. 

Let us consider the space of finite-support functions $C_{0}(\mathcal{S}(G))$ and the convolution product $\star_S\,\colon\,C_{0}(\mathcal{S}(G))\times C_{0}(\mathcal{S}(G)) \, \rightarrow C_{0}(\mathcal{S}(G))$, given as:
\begin{eqnarray}
&(f_1\star_{S}f_2)(\Gamma)= (f_1\star_{S}f_2)(\alpha,\beta, \gamma) = \sum_{\Gamma_1\in \mathcal{S}^{t_1(\Gamma)}}\nu_{(2)}^{t_1(\Gamma)}(\Gamma_1) f_1(\Gamma_1) f_2(\Gamma_1^{-1}\circ \Gamma) =\nonumber \\
& =\sum_{(\alpha_1,\gamma_1)\in G^{t(\alpha)}\times G^{t(\gamma)}} \nu^{t(\alpha)}(\alpha_1)\nu^{t(\gamma)}(\gamma_1)f_{1}(\alpha_1,\beta_1,\gamma_1)f_2((\alpha_1^{-1},\alpha_1 \circ \beta_1 \circ \gamma_1^{-1}, \gamma_1^{-1})\circ_V(\alpha,\beta,\gamma)) = \nonumber \\
& =\sum_{(\alpha_1,\gamma_1)\in G^{t(\alpha)}\times G^{t(\gamma)}}\nu^{t(\alpha)}(\alpha_1)\nu^{t(\gamma)}(\gamma_1) f_{1}(\alpha_1,\beta_1,\gamma_1)f_2(\alpha_1^{-1}\circ \alpha,\beta, \gamma_1^{-1}\circ \gamma) \,.
\end{eqnarray}
An involution operation $\ast$ is defined as follows
\begin{equation}
f^{\ast}(\Gamma)= \Delta_{(2)}^{-1}(\Gamma)\overline{f(\Gamma^{-1})}\,.
\end{equation} 
Let $\mathcal{H}_{S}=\mathcal{L}^2(\mathcal{S}(G), \mu_{(2)})$ be the Hilbert space of square-integrable functions on $\mathcal{S}(G)$ with respect to the measure $\mu_{(2)}$. The set $C_{0}(\mathcal{S}(G))$ is a dense subset of $\mathcal{H}_S$ and form an algebra under the convolution product. For any function $f\in C_{0}(\mathcal{S}(G))$ we can define the bounded operator $\mathbf{A}_f$ on the dense domain $C_{0}(\mathcal{S}(G))\subset \mathcal{H}_S$ as follows
\begin{equation}
(\mathbf{A}_f \psi)(\Gamma) = (f\star_S \psi)(\Gamma)\,.
\end{equation}
Since 
\begin{equation}
\mathbf{A}_f \mathbf{A}_g \psi = \mathbf{A}_{f\star_S g} \psi\,,\quad \mathbf{A}_f^{\dagger} = \mathbf{A}_{f^{\ast}}\,,
\end{equation}
the set $\Pi(C_0) = \left\lbrace \mathbf{A}_f\mid f\in C_0(\mathcal{S}(G)) \right\rbrace$ determines a $\star$-representation of the algebra $C_0(\mathcal{S}(G))$ supported on $\mathcal{H}_S$. 

Then, consider all the points $\Gamma = (\alpha,\beta,\gamma)\in \mathcal{S}(G)$ such that $\mu_{(2)}(\Gamma)\neq 0$. These are all the atoms of the measure $\mu_{(2)}$ and the characteristic functions $\delta_{\Gamma}$ form a basis of the algebra $C_{0}(\mathcal{S}(G))$ and of $\mathcal{H}_S$. Then, it is immediate to prove the following equality
\begin{equation}
\delta_{\Gamma} = \delta_{\mathbf{1}_{\beta}}\star_{S}\delta_{\Gamma},,
\end{equation}
so that the subset 
\begin{equation}
\Pi^{2}(C_0):=\left\lbrace f\star_S g \mid f,g\in C_0(\mathcal{S}(G))\right\rbrace
\end{equation}
is dense in $\mathcal{H}_S$. Collecting the above results we can conclude the following
\begin{proposition}
The set $C_0(\mathcal{S}(G))$ is a left Hilbert algebra within the Hilbert space $\mathcal{H}_S$ and the closure $\mathcal{V}(\mathcal{S})$ of the set $\Pi(C_0)$ with respect to the weak operator topology (or equivalently, the strong operator topology) is the associated left von-Neumann algebra.
\end{proposition}
We will call $\mathcal{V}(\mathcal{S})$ the Von Neumann symmetroid-algebra of the canonical symmetroid $\mathcal{S}(G)$. It is also possible to show, following the methods in \cite{non-commutative-1978,Hahn-1978}, that this algebra possesses a modular operator which is the multiplication operator by the function $\Delta_{(2)}$ and a modular involution, which is the operator $J_{(2)}\,\colon\,\mathcal{H}_{S}\,\rightarrow\,\mathcal{H}_S$ acting as follows
\begin{equation}
(J_{(2)}\psi)(\Gamma) = \overline{\psi(\Gamma^{-1})}\,.
\end{equation}

If we call $\Pi\,\colon\,C_0(\mathcal{S}(G))\,\rightarrow\,\mathcal{B}(\mathcal{H}_S)$ the representation which associates to every function $f\in C_0(\mathcal{S}(G))$ the corresponding bounded operator $\mathbf{A}_f$, we have just seen that it is a nondegenerate representation, but it will be highly reducible. The example of the groupoid of pairs $G(\Omega)$ over a finite set endowed with the counting measure will be considered in the rest of this section to illustrate the situation (in the finite connected case, indeed, every groupoid is the extension of a pair groupoid \cite{Ibort2019}). Additionally, this groupoid will play a special role in what follows because it is associated to the description of finite dimensional quantum systems in the Schwinger's picture of Quantum Mechanics.

Given a finite set $\Omega$ with cardinality $n$, whose elements will be denoted by $j= 1,\,2,\,\cdots, \,n$, the finite groupoid $G(\Omega)$ is made up of the pairs $(j,k)$ with $j,k=1,2,\cdots,n$. We endow this groupoid with the counting measure, each atom having the same measure, say $\mu(j,k)=1$, for every pair $(j,k)\in G(\Omega)$. With this measure, the modular function will be the trivial homomorphism, sending every pair to the value $1$.  Since the isotropy subgroupoid of the groupoid $G(\Omega)$ is trivial, the canonical symmetroid $\mathcal{S}(G(\Omega))$ reduces to the symmetroid $\tilde{\mathcal{S}}$ described in sec.\ref{canonical symmetroid} Let us remark, moreover, that this groupoid was firstly analysed in Ref.\cite{CIM-2019a}, where the concept of Schwinger's algebra of selective measurements was firstly analysed in the context of groupoids. The set $\mathcal{H}_S$ is isomorphic to $C_0(\tilde{\mathcal{S}})$ and both of them are finite dimensional, of dimension $N^2$ where $N=n^2$. Every function $f\in C_0(\tilde{\mathcal{S}})$ is represented by the collection of numbers $f((l,j),(k,m))$, and the convolution product and involution are defined as follows:
\begin{eqnarray}
&(f\star_S g)((l,j),(k,m)) = \sum_{r,s=1}^{n} f((l,r),(s,m)) g((r,j),(k,s)) \nonumber\\ 
&f^{\ast}((l,j),(k,m)) = \overline{f((j,l),(m,k))}\,,
\end{eqnarray}
where we have used the fact that $\Delta_{(2)}$ is the trivial function $\Delta_{(2)}((l,j),(k,m))=1$ for every $((l,j),(k,m))\in \tilde{\mathcal{S}}$. Therefore, we can identify this algebra with the tensor product algebra
\begin{equation}
C_{0}(\tilde{\mathcal{S}}) \cong M_n(\mathbb{C})\otimes M_n(\mathbb{C})\,,
\end{equation}
where $M_n(\mathbb{C})$ denotes the algebra on $n\times n$ matrix with complex entries.

The representation $\Pi\,\colon\,C_{0}(\tilde{\mathcal{S}})\,\rightarrow\,\mathcal{B}(\mathcal{H}_S)$ is supported on the Hilbert space $\mathcal{H}_S \cong \mathbb{C}^{N^2}$, so that the function $f\in C_0(\tilde{\mathcal{S}})$ is represented by the bounded operator
\begin{equation}
(\mathbf{A}_f \psi) ((l,j),(k,m)) = \sum_{r,s=1}^{n} f((l,r),(s,m)) \psi((r,j),(k,s))\,,
\end{equation}
and one can notice that the representation is highly reducible. Indeed, the subspace of functions which are constants along the fibers $\tilde{\mathcal{S}}^{(l,m)}$, i.e., the subspace $\mathcal{H}^{(l,m)}$ of functions $\psi = (t_1^*\tilde{\psi})$ which are the pullback of functions $\tilde{\psi}$ on $G(\Omega)$ via the projection $t_1$, are invariant under the action of the operators $\left\lbrace \mathbf{A}_{f} \right\rbrace$. Indeed, we have that
\begin{eqnarray}
&(\mathbf{A}_f (t_1^*\tilde{\psi}))((l,j),(k,m)) = f\star_S(t_1^*\tilde{\psi})((l,j),(k,m)) = \nonumber \\
&= \sum_{r,s=1}^n f((l,r),(s,m))(t_1^*\tilde{\psi})((r,j),(k,s)) = \sum_{r,s=1}^n f((l,r),(s,m))\tilde{\psi}(r,s)\,,
\end{eqnarray}
so that $(\mathbf{A}_f (t_1^*\tilde{\psi}))((l,j),(k,m))$ is constant along the fiber $\tilde{\mathcal{S}}^{(l,m)}$. This subset is isomorphic to the set of square-integrable functions on $G(\Omega)$, say $\mathcal{H}_G:=\mathcal{L}^2(G(\Omega))\cong \mathbb{C}^N$: The representation $\Pi(C_0)$ with support restricted to $\mathcal{H}^{(l,m)}\cong \mathcal{H}_G$ is irreducible since it is isomorphic to the representation of the algebra $M_n(\mathbb{C})\otimes M_n(\mathbb{C})$ on $M_n(\mathbb{C})\cong \mathbb{C}^N\cong \mathcal{H}_G$.   

The space $\mathcal{H}^{(l,m)}$ inherits another structure from the horizontal composition. Indeed, we can define the following product
\begin{equation}
((t_1^*\tilde{\psi_1})\star_H (t_1^*\tilde{\psi_2}))(\Gamma) = \sum_{\Gamma_1\circ_H\Gamma_2=\Gamma} (t_1^*\tilde{\psi_1})(\Gamma_1)(t_1^*\tilde{\psi_1})(\Gamma_2) = \sum_{\Gamma_1\circ_H\Gamma_2=\Gamma}\tilde{\psi}(t_1(\Gamma_1))\tilde{\psi(t_1(\Gamma_2))}\,,
\end{equation} 
which can be also rewritten as
\begin{equation}
((t_1^*\tilde{\psi_1})\star_H (t_1^*\tilde{\psi_2}))((l,j),(k,m)) = \sum_{r=1}^n \tilde{\psi}(l,r)\tilde{\psi(r,m)} = t_1^*(\tilde{\psi}_1\star \tilde{\psi}_2)((l,j),(k,m))\,.
\end{equation}
This means that the space $\mathcal{H}^{(l,m)}$ endowed with the product $\star_H$ is isomorphic as an algebra to the convolution algebra $(C_0(G(\Omega)), \star)$. Let us remark, however, that this definition of the horizontal convolution is adapted to the finite case. A different procedure should be thought for the continuum case, introducing a suitable family of measure as it is done for the convolution product of the groupoid-algebra $\mathcal{V}(G)$. 

\subsection{Dynamical maps}
In this section we will see how quantum dynamical maps can be described within Schwinger's picture of Quantum Mechanics. In order to avoid mathematical technicalities which at this stage would not help to clarify the idea behind this work, we will consider finite dimensional quantum systems, described in the Schwinger's picture by the groupoid $G(\Omega)$ with $\Omega$ a finite set. 

A dynamical map is a linear map $K \,\colon\,M_{n}(\mathbb{C})\,\rightarrow\,M_{n}(\mathbb{C})$\footnote{\begin{scriptsize}The case of map between quantum systems with different dimensions, say $n_1 < n_2$, can be dealt in this frame by considering an immersion the algebra $M_{n_2}(\mathbb{C})$ within $M_{n_1}(\mathbb{C})$ and considering the extension $\tilde{K}$ by zeroes of the dynamical map $K\,\colon\,M_{n_1}(\mathbb{C})\,\rightarrow\,M_{n_2}(\mathbb{C})$.\end{scriptsize}} sending positive operators to positive operators and preserving the unit. In this way, the dual map $K^{\ast}$ will send quantum states to quantum states. For physical purposes\cite{Lindblad-1976,GKS-1976}, the special class of completely positive and unital maps play a distinguished role and are usually called channels. A dynamical map $K\,\colon\,M_{n}(\mathbb{C})\,\rightarrow\,M_{n}(\mathbb{C})$ is said to be completely positive if it is positive and the map 
\begin{equation}
K\otimes \mathbb{I}_M = K_M \,\colon\,M_{n}(\mathbb{C})\otimes M_{M}(\mathbb{C})\,\rightarrow\,M_{n}(\mathbb{C})\otimes M_{M}(\mathbb{C})
\end{equation}
is positive for every $M\in \mathbb{N}$. 

In Schwinger's picture of Quantum Mechanics, a state of a finite dimensional quantum system is described by a function of positive-type on the groupoid as in Eq.(\ref{positive_definite_function}). Therefore, a dynamical map is a linear map on the space of functions $C_0(G)$ on the groupoid which preserves the positivity condition. A class of dynamical maps which are naturally defined in this groupoidal frame is given in terms of flat bisections of the canonical symmetroid $\tilde{\mathcal{S}}$. As we have described in Sec.\ref{sec:flat_bisections}, a flat bisection $\mathbf{b}\subset \tilde{\mathcal{S}}$ of the canonical symmetroid determines a functor $\phi_{b}$ whose action on morphisms of the groupoid $G(\Omega)$ is expressed as in Eq.(\ref{flat_functor}), i.e., 
\begin{equation}
\phi_{b}(j,k) = b_s(j)\circ (j,k)\circ b_s(k)^{-1} = (b_s(j),b_s(k))\,,
\end{equation}
where $b_s\,\colon\, \Omega\,\rightarrow\,\Omega$ is a bijective map on the finite set $\Omega$ (indeed, the bisections of the pair groupoid $G(\Omega)$ are the graphs of bijective maps, see \cite{CDIMS-2021}). 

Since the functor $\varphi\,\colon\,G(\Omega)\,\rightarrow\,G(\Omega)$ is an automorphism of the groupoid $G(\Omega)$, it induces, via pullback, the map $\phi_b^{*}\,\colon\,C_0(G(\Omega))\,\rightarrow\,C_0(G(\Omega))$ on the space of functions on $G(\Omega)$. Since a functor preserves the composition law in the groupoid $G(\Omega)$, the map $\phi_b^*$ is a linear map which preserves the positivity. Indeed, given the flat bisection $\mathbf{b}\subset \tilde{\mathcal{S}}$, the corresponding functor $\phi_b$, and a positive definite function $\varphi_\rho$ on $G(\Omega)$, we have that:
\begin{eqnarray}
&\sum_{j,k=1}^N \xi_j \overline{\xi}_k \phi^*(\varphi_{\rho})(\alpha_j\circ \alpha_k^{-1}) = \nonumber \\
&= \sum_{j,k=1}^N \xi_j \overline{\xi}_k \varphi_{\rho}(\phi_b(\alpha_j\circ \alpha_k^{-1})) = \sum_{j,k=1}^N \xi_j \overline{\xi}_k \varphi_{\rho}(\phi_b(\alpha_j)\circ \phi_b(\alpha_k)^{-1}))\,,
\end{eqnarray}    
and since $\varphi_{\rho}$ is positive-type, the same holds for $\phi_b^*(\varphi_{\rho})$. It is clear that linear combination of these maps with positive coefficients are dynamical maps, too. 

Let $\Omega_M$ be a finite set of cardinality $M$ and $G(\Omega_M)\times G(\Omega)\rightrightarrows \Omega_M\times \Omega$ the direct product of the pair groupoids over $\Omega_M$ and $\Omega$, respectively (see \cite{Ibort2019,CDIM-2020b} for more details on composition). The functor $\phi_b$ on $G(\Omega)$ can then be extended to the functor 
\begin{equation}
\phi_b\times \mathrm{id}_M = \phi_b^M\,\colon\,G(\Omega_M)\times G(\Omega)\,\rightarrow\,G(\Omega_M)\times G(\Omega)\,,
\end{equation}
i.e., the product of the functor $\phi_b$ with the identity functor $\mathrm{id}_M$ on the groupoid $G(\Omega_M)$. Now, since the map $\phi_b^M$ preserves the composition law of the product groupoid, the pullback $(\phi_b^M)^*$ maps positive definite functions to positive definite functions, and this is valid for every $M\in \mathbb{N}$. From this we conclude that the dynamical maps $\phi^*_b$ determined by flat bisections $\mathbf{b}$ are completely positive. 

However, the symmetroid-algebra allows us to look at the description of dynamical maps from a different perspective. Indeed, let us consider the transformation $\Gamma=((l,j),(k,m))$ and the characteristic function $\delta_{\Gamma}$ which takes the value one only when its argument is equal to $\Gamma$ and zero otherwise. A straightforward computation shows that
\begin{eqnarray}
&(\delta_{\Gamma}\star_S t_1^*\tilde{\psi}) ((l,j'),(k',m)) = \sum_{r,s} \delta_{\Gamma}((l,r),(s,m))\tilde{\psi}(r,s) = \nonumber\\
&=\sum_{r,s} \delta_{(l,j)}(l,r)\tilde{\psi}(r,s)\delta^{\ast}_{(m,k)}(s,m) = ( \delta_{(l,j)}\star \tilde{\psi} \star \delta^{\ast}_{(m,k)} )(l,m)\label{eq_3.64}\,,
\end{eqnarray}   
so that the action by $\star_S$-convolution of a characteristic function on the symmetroid is expressed in terms of the characteristic functions $\delta_{(l,j)}$ and $\delta^*_{(k,m)}$ on the groupoid $G(\Omega)$ acting by $\star$-convolution on the right and on the left, respectively. These functions form a basis of the vector space $C_0(\tilde{\mathcal{S}})$, so that any element of the symmetroid-algebra is represented as a linear combination of them. This representation corresponds to the $A$-matrix introduced by Sudarshan et al \cite{Sudarshan1961}. On the other hand, there is another way to represent the action of the operator $\delta_{\Gamma}$. Indeed, we can repeat the computation in Eq.(\ref{eq_3.64}) in a different way:  
\begin{eqnarray}
&(\delta_{\Gamma}\star_S t_1^*\tilde{\psi}) ((l,j'),(k',m)) = \sum_{r,s} \delta_{\Gamma}((l,r),(s,m))\tilde{\psi}(r,s) = \nonumber\\
&=\sum_{r,s}\delta_{(l,m)}(l,m)\tilde{\psi}(r,s)\delta_{(j,k)}(r,s) = \delta_{(l,m)}(l,m)\left\langle \delta_{(j,k)} , \tilde{\psi} \right\rangle_G\,.
\end{eqnarray}
This second representation is known as the $B$-matrix representation \cite{Sudarshan1961}.

Now, dynamical maps are described by those functions $f_{K}$ which can be expanded as follows (the tensor product symbol is adapted to the $A$-matrix decomposition):
\begin{equation}\label{symmetroid_dynamical_maps}
f_K((l,j),(k,m)) = \sum_{p} V_p\otimes V_p^*((l,j),(k,m)) = V_p(l,j)V^{\ast}_p(k,m)\,,
\end{equation}    
where the number of functions $V_p$ can be at most $n^2$ and $f_K$ is unital only if
\begin{equation}
\sum_p V_p\star V_p^{\ast} = \chi_{\Omega}\,.
\end{equation}
These maps are all completely positive, due to Choi's theorem \cite{choi1975completely} and satisfy the following property: for any $M\in\mathbb{N}$ and every complex vector $\left\lbrace \xi_j \right\rbrace_{j=1,2,\cdots\,M}$  
\begin{equation}
\sum_{j,k=1}^M \overline{\xi}_k\xi_j f_K(\Gamma_j\circ_H \Gamma_k^{-H}  ) \geq 0
\end{equation}
for every $M$-ple of transformations $\left\lbrace \Gamma_j \right\rbrace$. We will call functions obeying this condition flat positive semidefinite functions. In general these functions are not of positive-type, since it holds that
\begin{equation}
\sum_{j,k=1}^M \overline{\xi}_k\xi_j f_K(\Gamma_j\circ_V \Gamma_k^{-1}  )= \sum_p \sum_{j,k=1}^M V_p(y_j,y_k)\overline{V_p(x_j,x_k)}\,,
\end{equation}
and it does not need to be positive. However, if $V_p$ is the function associated with a Dirac-Schwinger-Feynman state on $G(\Omega)$ \cite{CDIMSZ-2021b}, i.e., it satisfies
\begin{equation}\label{DSF-states}
V_p(j,k)V_p(k,l)=V_p(j,l)\,,\quad V_p^{\ast} = V_p
\end{equation}
the corresponding $f_K= \sum_{p} V_p\otimes V_p^* $ is both positive and flat positive semidefinite. Moreover this map will be unital iff 
\begin{equation}
\sum_p V_p = \sum_p V_p\star V_p = \sum_p V_p \star V_p^{\ast} = \chi_{\Omega}\,.
\end{equation}

We conclude the section with two examples of dynamical maps illustrating the procedure described in this section. The first example consists of the map associated with the functor $\phi_b$ corresponding to a flat bisection $\mathbf{b}\subset \tilde{\mathcal{S}}$. Let $b\subset G(\Omega)$ be the subset 
\begin{equation}
b=\left\lbrace (j+1, j)\mid j\in \Omega \right\rbrace\,,
\end{equation}
and let $\mathbf{b}\subset \tilde{\mathcal{S}}$ be the flat bisection
\begin{equation}
\mathbf{b} = \left\lbrace ((j+1,j),(k,k+1))\mid (j,k)\in G(\Omega) \right\rbrace\,.
\end{equation}
The dynamical map $\phi_b^*$ acts on a function $\psi\in C_0(G(\Omega))$ as follows
\begin{equation}
(\phi_b^*\psi)(j,k) = \psi (\phi_b(j,k)) = \psi(j+1, k+1)\,.
\end{equation}
Let $\chi_b$ the characteristic function on the set $b$ and $\tilde{\chi}_{\mathbf{b}}$ be the characteristic function on the flat bisection $\mathbf{b}$. Then, the action of the map $\phi_b^*$ can be alternatively represented in terms of the convolution product $\star_S$:
\begin{equation}
(\tilde{\chi}_{\mathbf{b}}\star_St_1^*\tilde{\psi})((l,p),(q,m))= \sum_{r,s}\chi_b(l,r) \tilde{\psi}(r,s) \chi_b^*(s,m) = \chi_b\star \tilde{\psi}\star \chi_b^*\,,
\end{equation}
where the characteristic function $\chi_b$ determines the following unitary operator $U_b$ on the space $\mathcal{H}_G$ 
\begin{equation}
U_b\tilde{\psi} = \chi_b\star \tilde{\psi}\,.
\end{equation}
Therefore, the functor $\phi_b$ generates the action by conjugation by the unitary operator $U_b$ on the algebra $M_n(\mathbb{C})$, confirming the fact that functors define automorphisms of the groupoid and, consequently, of the groupoid-algebra.

The second example is given by the map generated by functions $V_p$ as in Eq.(\ref{DSF-states}). In particular, we can choose the following family of functions $\left\lbrace V_l \right\rbrace_{l=1,2,\cdots,n}$:
\begin{equation}
V_l(j,k) = \frac{1}{n}\mathrm{e}^{i\frac{2\pi}{n}l(j-k)}\,.
\end{equation}
Then, the dynamical map $f_K = \sum_{l=1}^n V_l\otimes V_l$ acts as follows:
\begin{eqnarray}
&(f_K\star_S t_1^*\tilde{\psi})((p,j),(k,m))= \sum_{l=1}^n\sum_{r,s=1}^n \frac{1}{n}\mathrm{e}^{i\frac{2\pi}{n}l(p-r)}\tilde{\psi}(r,s)\frac{1}{n}\mathrm{e}^{i\frac{2\pi}{n}l(s-m)} = \nonumber \\
&=\sum_{l=1}^n \left\langle \tilde{\psi} \right\rangle_l \frac{1}{n}\mathrm{e}^{i\frac{2\pi}{n}l(p-m)}\,,
\end{eqnarray}
where 
\begin{equation}
\left\langle \tilde{\psi} \right\rangle_l = \sum_{r,s=1}^n \frac{1}{\sqrt{n}}\mathrm{e}^{-i\frac{2\pi}{n}lr}\tilde{\psi}(r,s)\frac{1}{\sqrt{n}}\mathrm{e}^{i\frac{2\pi}{n}ls}\,,
\end{equation}
are the expectation values of the matrix associated with the function $\tilde{\psi}$ on the vectors $| v_l \rangle \in \mathbb{C}^n$ with components
\begin{equation}
| v_l \rangle = \left\lbrace \frac{1}{\sqrt{n}}\mathrm{e}^{i\frac{2\pi}{n}ls} \right\rbrace_{s=1,2,\cdots,n}\,.
\end{equation}
The family of vectors $\left\lbrace v_l \right\rbrace_{l=1,2,\cdots,n}$ form an orthonormal basis of the Hilbert space $\mathbb{C}^n$ and the dynamical map $f_K$ associates with each $\tilde{\psi}$ its tomogram with respect to this orthonormal basis, describing a decoherent dynamics, in the sense that starting from a generic quantum states it gives as a result a state on the maximal Abelian subalgebra determined by the projectors on the vectors of the given orthonormal basis. 

\section{Conclusions}
In this paper the description of dynamical maps beyond unitary ones is addressed in the groupoid approach to Quantum Mechanics. The departure point is the canonical symmetroid $\tilde{\mathcal{S}}$ associated with the pair groupoid $G(\Omega)$ on a finite set $\Omega$ of cardinality $n$. After the definition of a suitable Haar measure on the canonical symmetroid which is induced from a Haar measure on the groupoid $G(\Omega)$, the von-Neumann symmetroid-algebra $\mathcal{V}(\tilde{\mathcal{S}})$ is introduced as the closure of a set of operators acting on the Hilbert space of square-integrable functions on the symmetroid. The action of these operators can be restricted to a Hilbert subspace which is isomorphic to the space $M_n(\mathbb{C})$, and a subset of these operators is proven to determine completely positive maps on $M_n(\mathbb{C})$. In particular, it is shown that these functions satisfy a positivity condition which we have called flat positive semidefiniteness, which provides an analogue for the channel-state duality in this groupoidal approach. Moreover, the descriptions in terms of $A$-matrix and $B$-matrix are also obtained: they have been related to two different ways in which a basis of the space of functions on the canonical symmetroid $\tilde{\mathcal{S}}$ can be expressed as the product of functions on the groupoid $G(\Omega)$.

In this work, we focused on the finite-dimensional case in order to convey the main idea without burdening the discussion with the functional analytic technicalities characteristic of the infinite-dimensional case. However, the construction presented here can and will be extended to the infinite-dimensional setting in future, forthcoming works where we will also investigate possible connections with the results given in \cite{Holevo2011,Stormer2015}.

\section*{Acknowledgments}
The authors acknowledge financial support from the Spanish Ministry of Economy and Competitiveness, through the Severo Ochoa Programme for Centres of Excellence in RD (SEV-2015/0554), the MINECO research project  PID2020-117477GB-I00,  and Comunidad de Madrid project QUITEMAD++, S2018/TCS-A4342.
GM would like to thank partial financial support provided by the Santander/UC3M Excellence  Chair Program 2019/2020, and he is also a member of the Gruppo Nazionale di Fisica Matematica (INDAM), Italy. 
FDC acknowledges support from the CONEX-Plus programme funded by Universidad Carlos III de Madrid and the European Union's Horizon 2020 research and innovation programme under the Marie Sklodowska-Curie grant agreement No. 801538. 
FMC acknowledges that the work has been supported by the Madrid Government (Comunidad de Madrid-Spain) under the Multiannual Agreement with UC3M in the line of ``Research Funds for Beatriz Galindo Fellowships'' (C$\setminus$\&QIG-BG-CM-UC3M), and in the context of the V PRICIT (Regional Programme of Research and Technological Innovation).  

\bibliography{channel-state_duality}
\bibliographystyle{plain}

\end{document}